\documentclass[final,5p,times]{elsarticle}

\usepackage[numbers]{natbib}
\usepackage{tabularx}
\usepackage{float}
\usepackage{graphicx}
\usepackage{url}
\usepackage{xcolor}
\usepackage{tcolorbox}
\usepackage{amsmath}
\usepackage{amsfonts}
\usepackage{microtype}
\usepackage{multirow}
\usepackage{booktabs}
\usepackage{listings}
\usepackage{pifont}
\usepackage{makecell}
\usepackage{placeins}
\usepackage{dblfloatfix}
\usepackage{subcaption}
\usepackage{nicematrix}
\usepackage{bm}
\usepackage{caption}
\usepackage{enumitem}
\usepackage{threeparttable}
\usepackage{orcidlink}
\usepackage[ruled,vlined,linesnumbered]{algorithm2e}

\definecolor{verylightgray}{rgb}{.97,.97,.97}
\journal{}

\begin{document}

\begin{frontmatter}
	
	\title{Security Attack and Defense Strategies for Autonomous Agent Frameworks: A Layered Review with OpenClaw as a Case Study}

    \author[NTU]{Luyao Xu\orcidlink{0009-0006-6983-7134}}	
\ead{lxu62519@gmail.com}    	
\author[NTU,NJU]{Xiang Chen\orcidlink{0000-0002-1180-3891}\corref{mycorrespondingauthor}}        
\cortext[mycorrespondingauthor]{Corresponding author}	
\ead{xchencs@ntu.edu.cn}			    			
\address[NTU]{School of Artificial Intelligence and Computer Science, Nantong University, Nantong, China}
\address[NJU]{State Key Lab. for Novel Software Technology,
Nanjing University, Nanjing, China}
\begin{abstract}

Autonomous agent frameworks built upon large language models (LLMs) are evolving into complex, tool-integrated, and continuously operating systems, introducing security risks beyond traditional prompt-level vulnerabilities. As this paradigm is still at an early stage of development, a timely and systematic understanding of its security implications is increasingly important. Although a growing body of work has examined different attack surfaces and defense problems in agent systems, existing studies remain scattered across individual aspects of agent security, and there is still a lack of a layered review on this topic. To address this gap, this survey presents a layered review of security risks and defense strategies in autonomous agent frameworks, with OpenClaw as a case study. We organize the analysis into four security-relevant layers: the context and instruction layer, the tool and action layer, the state and persistence layer, and the ecosystem and automation layer. For each layer, we summarize its functional role, representative security risks, and corresponding defense strategies. Based on this layered analysis, we further identify that threats in autonomous agent frameworks may propagate across layers, from manipulated inputs to unsafe actions, persistent state contamination, and broader ecosystem-level impact. Finally, we highlight potential key challenges, including research imbalance across layers, the lack of long-horizon evaluation, and weak ecosystem trust models, and outline future directions toward more systematic and integrated defenses.

\end{abstract}


\end{frontmatter}

\section{Introduction}
\label{sec:intro}
Autonomous agent frameworks are emerging as a new class of LLM-based systems that extend far beyond conventional chat assistants. By enabling large language models (LLMs) to invoke external tools, maintain persistent state, incorporate extensible skills, and execute long-horizon tasks with limited human intervention, these frameworks provide the infrastructure for building agents capable of autonomous planning, decision-making, and action. Previous studies on reasoning-and-acting, tool use,
and generative agents have already shown how LLMs can be coupled with planning,
memory, and external interfaces to support more autonomous behavior
~\cite{yao2023react,schick2023toolformer,park2023generative,wang2024survey}.
Recent real-world frameworks further operationalize these ideas into deployable
agent infrastructures. Among them, OpenClaw is a representative case in which
system prompts are dynamically assembled from tools, skills, runtime metadata, and workspace files, while agent capabilities are extended through typed tools, workspace context, and automation mechanisms such as heartbeat and cron~\cite{openclaw_systemprompt,openclaw_context,openclaw_tools,openclaw_skills,openclaw_workspace,openclaw_heartbeat,openclaw_cron}.

These capabilities substantially enlarge the security boundary of agent systems.
In traditional LLM applications, security concerns are often discussed at the level
of unsafe or manipulated text generation. In autonomous agent frameworks,
however, attacks can propagate from contextual inputs to concrete tool execution,
persistent state, and external ecosystems. As a result, the threat landscape is no
longer limited to prompt-level manipulation, but extends to unsafe actions,
state contamination, ecosystem and supply-chain compromise, and automation-driven
abuse. OpenClaw's own security guidance reflects this broader risk model by
explicitly emphasizing delegated tool authority, mixed-trust inputs, shared state,
and the importance of runtime controls such as policy and sandboxing
~\cite{openclaw_security,openclaw_workspace,openclaw_tools}.

This broader threat model has already motivated a growing body of security research, but the literature remains fragmented. Existing studies have investigated agent security from different perspectives. Some studies focus on benchmark-driven robustness evaluation, such as AgentDojo and Agent Security Bench (ASB)~\cite{agentdojo,zhang2025asb}, which assess agent behavior under adversarial instructions and tool-use scenarios. Other works study framework-specific risks, such as PASB~\cite{wang2026pasb}, which evaluates security issues in OpenClaw-centered agent settings. Recent studies further examine specific threat mechanisms, including stealthy token exhaustion, guidance injection, and defense-oriented architectural redesign~\cite{clawdrain,liu2026trojanswhisper,openclaw_defensible_design}.
Although these studies provide valuable insights, they remain scattered across different attack surfaces, system assumptions, and evaluation settings. This fragmentation makes it difficult for researchers and practitioners to understand the current landscape, compare security threats and defenses across the core layers of autonomous agent frameworks, and identify open challenges. Therefore, a comprehensive survey is needed to systematically organize existing work, analyze emerging trends, and highlight future research directions for security in autonomous agent frameworks.

To address this gap, we present a layered review of security risks and defense mechanisms in autonomous agent frameworks, with OpenClaw as a case study. We organize the analysis around four security-relevant layers: the context and instruction layer, the tool and action layer, the state and persistence layer, and the ecosystem and automation layer. For each layer, we systematically review its functional role, representative threat models, attack techniques, and defense strategies, thereby clarifying how existing studies address security risks at different architectural levels. Based on this layered analysis, we further identify that security risks in agent systems are inherently cross-layer: attacks can originate from manipulated inputs, propagate through tool invocation and action execution, persist in memory or workspace state, and eventually amplify across external environments and automation mechanisms. This end-to-end attack lifecycle reveals a fundamental mismatch between the evolving threat landscape and existing defense strategies, many of which are still designed for isolated components rather than cross-layer agent systems.

In summary, the main contributions of this survey are threefold: 
\begin{itemize}
  \item We provide a unified layered taxonomy for analyzing agent security.
  \item We systematically synthesize existing threat and defense studies across different architectural levels.
	\item We highlight cross-layer threat propagation together with key research gaps in current agent security research.
\end{itemize}

The remainder of this survey is organized as follows. Section~\ref{sec:related_surveys} reviews related surveys and security studies and positions this work within the existing literature. Section~\ref{sec:Framework Overview} introduces the framework overview of autonomous agent frameworks. Section~\ref{sec:Layered Attacks and Defenses} presents the layered analysis of attacks and defenses. Section~\ref{sec:Cross-Layer Attacks and Defenses} compares representative studies and discusses cross-layer attack propagation and defense limitations. Section~\ref{sec:Open Problems and Future Directions} summarizes open problems and future research directions, and Section~\ref{sec:conclusion} concludes the survey.

\section{Related Surveys and Security Studies}
\label{sec:related_surveys}

Several lines of prior work are related to this survey, spanning general LLM safety reviews, prompt injection analyses, and agent-specific security studies. We discuss each category below, highlighting how they contribute to the broader landscape and where gaps remain that motivate our survey.

\textbf{General LLM safety surveys.} A first category of related work reviews security and safety risks of large language models at a broad level. For example, Dong et al.~\cite{safeguarding_llms} provide a comprehensive survey of guardrails and safeguard mechanisms for LLMs, covering alignment, toxicity, jailbreaking, and output control. Wang et al.~\cite{wang2024survey} survey LLM-based autonomous agents with a focus on architecture, capabilities, and applications, touching briefly on safety and trustworthiness. While these surveys offer valuable overviews, they are not organized around the security challenges unique to autonomous agent frameworks. In particular, they do not systematically examine how tool execution, persistent state, and ecosystem integrations create new and interconnected attack surfaces that go beyond prompt-level or alignment-level risks.

\textbf{Prompt injection surveys.} A second category provides in-depth analysis of prompt injection attacks, which represent the most extensively studied threat to LLM-based systems. Geng et al.~\cite{prompt_injection_survey} survey attack methods, root causes, and defense strategies for prompt injection, offering a detailed taxonomy of this specific attack surface. However, prompt injection surveys typically focus on the context and instruction layer and do not extend their analysis to downstream consequences such as unsafe tool invocation, state contamination, or ecosystem-level propagation. As a result, they capture only one dimension of the broader threat landscape facing autonomous agent frameworks.

\textbf{Agent-specific security studies.} A third and most closely related category includes works that directly examine security risks in LLM-based agent systems. Zhang et al.~\cite{zhang2025asb} propose Agent Security Bench (ASB), which formalizes and benchmarks attacks and defenses across multiple agent scenarios. Ferrag et al.~\cite{ferrag2026agents} analyze threats ranging from prompt injection to protocol exploits in LLM-powered agent workflows. Ruan et al.~\cite{toolemu} develop ToolEmu, an LM-emulated sandbox for identifying tool-use risks before deployment. Hua et al.~\cite{trustagent} introduce TrustAgent, a safety framework based on pre-planning and post-action safety checks. While these works directly address agent security, they are primarily oriented toward benchmarking, evaluation, or single-mechanism analysis rather than providing a layered survey that synthesizes attacks and defenses across different architectural levels and examines cross-layer threat propagation.

\begin{table*}[htbp]
\centering
\caption{Feature comparison of our survey with respect to previous related surveys and security studies.}
\label{tab:related_comparison}
\resizebox{\textwidth}{!}{
\begin{tabular}{cccccccc}
\toprule
\textbf{Study} & \textbf{Primary Focus} & \textbf{Scope} & \textbf{Organizing Principle} & \textbf{Layered} & \textbf{Cross-layer} & \textbf{Case Study} & \textbf{Ecosystem} \\
\midrule
Dong et al.~\cite{safeguarding_llms} & LLM safety and guardrails & LLM-level & Threat taxonomy & \ding{55} & \ding{55} & \ding{55} & Partial \\
Wang et al.~\cite{wang2024survey} & LLM autonomous agents & Agent-level & Capability taxonomy & \ding{55} & \ding{55} & \ding{55} & \ding{55} \\
Geng et al.~\cite{prompt_injection_survey} & Prompt injection & Prompt-level & Attack-defense cat. & \ding{55} & \ding{55} & \ding{55} & \ding{55} \\
Zhang et al.~\cite{zhang2025asb} & Agent security benchmark & Agent-level & Benchmark-driven & Partial & Partial & \ding{55} & \ding{55} \\
Ferrag et al.~\cite{ferrag2026agents} & Agent workflow threats & Workflow-level & Threat classification & Partial & Partial & \ding{55} & Partial \\
Ruan et al.~\cite{toolemu} & Tool-use safety evaluation & Tool-level & Emulation-based & Partial & Partial & \ding{55} & \ding{55} \\
Hua et al.~\cite{trustagent} & Agent safety framework & Agent-level & Constitution-based & Partial & Partial & \ding{55} & \ding{55} \\
\textbf{Our Survey} & \textbf{Agent framework security} & \textbf{Framework-level} & \textbf{Four-layer taxonomy} & \ding{51} & \ding{51} & \ding{51} & \ding{51} \\
\bottomrule
\end{tabular}
}
\end{table*}

Table~\ref{tab:related_comparison} provides a structured comparison across several key dimensions. Three observations emerge from this comparison. First, existing general surveys cover LLM safety or agent capabilities broadly but do not organize their analysis around the security-relevant architectural layers of autonomous agent frameworks. Second, prompt injection surveys offer depth on context-layer threats but do not address how those threats propagate into tool execution, persistent state, or ecosystem-level impact. Third, agent-specific security studies contribute important benchmarks and evaluation tools but remain focused on individual mechanisms or attack types rather than providing a cross-layer literature synthesis.
In contrast, our survey addresses these gaps by centering its analysis on autonomous agent framework security, adopting a four-layer taxonomy as the organizing principle, examining cross-layer attack propagation and defense mismatches, and using OpenClaw as a concrete case study to demonstrate the practical value of the layered perspective. In this sense, the present work is positioned not as another general survey of LLM safety or autonomous agents, but as a structured security analysis that introduces a new organizing principle and analytical granularity for understanding threats and defenses in agent frameworks.

\section{Framework Overview}
\label{sec:Framework Overview}
\begin{figure*}[htbp]
    \centering
    \includegraphics[width=\textwidth]{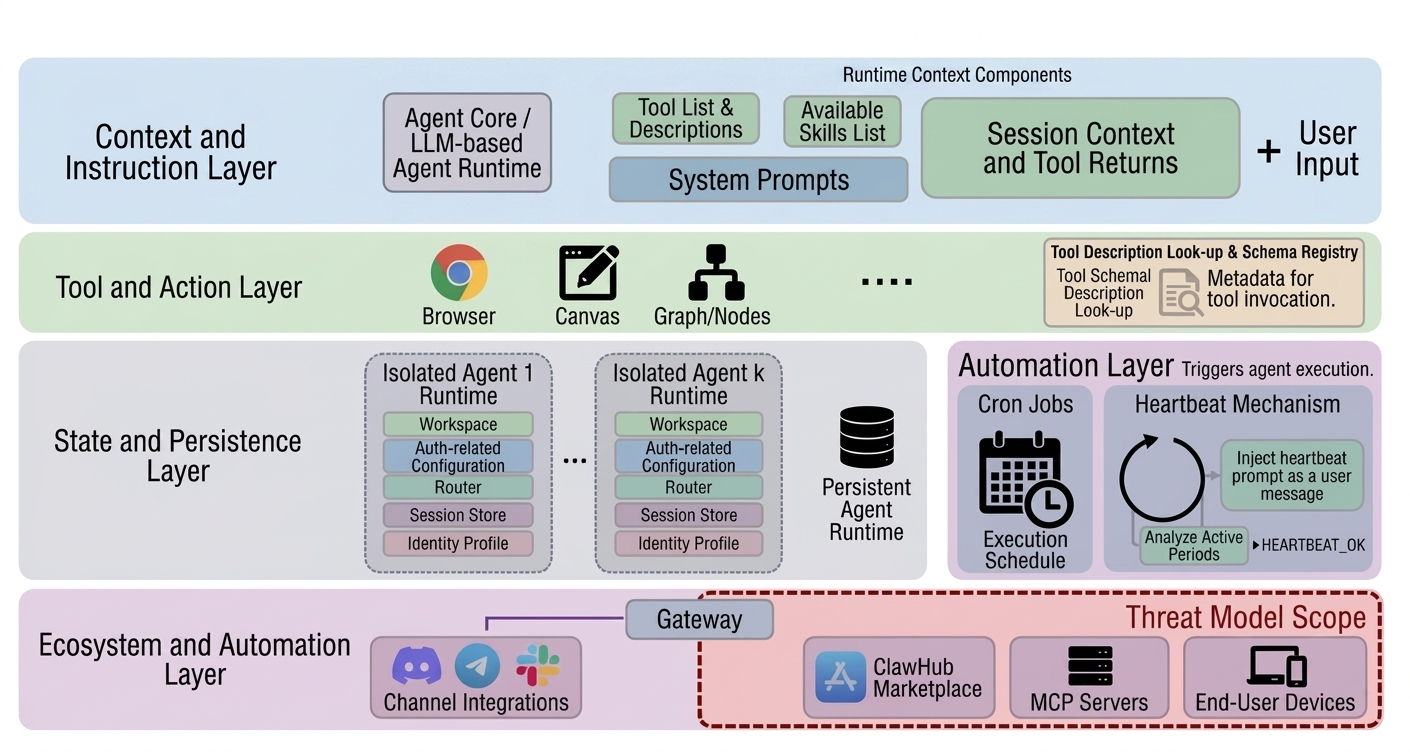}
    \caption{Layered Architecture of Autonomous Agent Frameworks: An Illustration Using OpenClaw as a Case Study.}
    \label{fig:framework_overview}
\end{figure*}

Autonomous agent frameworks are more than simple interfaces wrapped around large language models. In contrast to conventional chat-oriented systems, they combine contextual reasoning, executable tool use, persistent runtime state, and extensible ecosystem interfaces into a unified operational framework~\cite{yao2023react,schick2023toolformer,park2023generative,wang2024survey}.
From a security perspective, these capabilities define the main surfaces through which agent behavior can be shaped, extended, and exploited. To systematically analyze these surfaces, we propose a four-layer security taxonomy for autonomous agent frameworks, as summarized in Figure~\ref{fig:framework_overview}. This taxonomy organizes agent security around four closely related layers: the context and instruction layer, the tool and action layer, the state and persistence layer, and the ecosystem and automation layer. Taken together, these four layers provide a concise architectural view of autonomous agent frameworks and establish the structural basis for the security analysis developed in the following sections, where attacks and defenses are examined according to how they affect contextual reasoning, executable actions, persistent state, and ecosystem-level extension. In the rest of this section, we briefly introduce each layer.

\textbf{The context and instruction layer} determines how model-visible input is constructed during agent execution. 
Rather than relying solely on user prompts, modern agent frameworks assemble a composite context that integrates system prompts, external content, auxiliary files, skill descriptions, and tool returns ~\cite{wang2024survey,agentdojo}. In OpenClaw, this process is explicitly realized through runtime prompt construction and the injection of contextual artifacts, such as tool lists, available skills, bootstrap files, and session-level information~\cite{openclaw_systemprompt,openclaw_skills}.

\textbf{The tool and action layer} captures what the agent can do beyond text generation. In autonomous agent frameworks, model outputs are routinely translated into executable actions through typed tools, including file operations, browser control, web access, messaging, and device interaction ~\cite{schick2023toolformer,wang2024survey}. OpenClaw similarly exposes actions through its tool interfaces, where capabilities beyond text generation are implemented as tools rather than plain language outputs ~\cite{openclaw_tools}.

\textbf{The state and persistence layer} represents what the system preserves over time within the agent itself. 
Unlike single-turn assistants, agent frameworks retain runtime traces such as session history, workspace artifacts, identity-related settings, and other persistent configuration elements  ~\cite{park2023generative,wang2024survey}. In OpenClaw, such persistence is reflected in agent-specific workspaces, session stores, identity or profile settings, and related runtime state ~\cite{openclaw_workspace}.

\textbf{The ecosystem and automation layer} describes how the framework is extended and connected to external environments. Skills, plugins, channel integrations, marketplaces, and external services expand the functional scope of the agent~\cite{wang2024survey}. In OpenClaw, this layer plays a critical role in enabling integration with external capabilities, providing access to skills, services, and platform-level functionalities through a unified interface~\cite{openclaw_skills,openclaw_heartbeat,openclaw_cron,openclaw_threat_model}. Such integrations allow the agent to extend its functionality beyond the core system and operate across diverse external environments.

Previous agent surveys typically organize their analysis around capability modules such as planning, memory, tool use, and multi-agent coordination~\cite{wang2024survey}. While effective for characterizing agent functionality, such taxonomies do not directly capture how security threats propagate through an agent system. Our four-layer decomposition is instead motivated by the lifecycle of an attack: adversarial input first enters through the context layer, is then translated into concrete actions at the tool layer, may become persistent through the state layer, and can ultimately propagate across system boundaries via the ecosystem layer. This progression from entry to execution to persistence to amplification reflects the causal structure through which a single injected instruction can escalate into a systemic compromise. The chain-like relationship among the four layers also provides the structural foundation for the cross-layer propagation analysis presented in Section~\ref{sec:cross_layer_attack_propagation}, where we examine how vulnerabilities in one layer enable or amplify threats in adjacent layers.

\section{Layered Attacks and Defenses }
\label{sec:Layered Attacks and Defenses}
The following analysis adopts a layer-wise perspective to systematically examine security risks in autonomous agent frameworks. Rather than treating attacks in isolation, we focus on a set of security-relevant components that directly influence how agents construct context, execute actions, maintain state, and interact with external environments. These components serve as the analytical anchors for identifying attack surfaces and understanding defense mechanisms across layers. For each layer, we follow a consistent analytical framework by first clarifying its functional role and attack surface, then summarizing representative attacks and defenses, and finally discussing the main attack-defense gaps, limitations, and cross-layer implications.

\subsection{Context and Instruction Layer}
\label{sec:context_instruction_layer}

\textbf{Layer Role.}
The context and instruction layer governs how information is presented to the model at each reasoning step. It serves as the immediate input channel through which the agent receives task instructions, environmental information, and intermediate feedback needed for decision making~\cite{yao2023react,agentdojo}.

\textbf{Attack Surface.}
Security risks in this layer arise from the mixture of trusted control instructions (e.g., system prompts) and untrusted data (e.g., retrieved content or tool outputs). Because these elements are concatenated into a single context window, the model often lacks a reliable mechanism to distinguish control signals from data, making it vulnerable to manipulation at the input level~\cite{prompt_injection_survey,agentdojo}.

\textbf{Attacks.}
Attacks on the context and instruction layer can be organized according to how adversarial content enters and manipulates the agent's context.

First, attacks may target the direct user-agent instruction channel. Direct prompt injection attempts to override system-level or developer-level instructions through malicious user inputs, exploiting the model's difficulty in separating user requests from higher-priority control constraints~\cite{prompt_injection}.

Second, attacks may enter through external or retrieved content. Indirect prompt injection embeds adversarial instructions within web pages, documents, emails, or other external sources that are later incorporated into the agent's context. In this case, untrusted data is transformed into an implicit instruction source, allowing attackers to influence the agent without directly interacting with it~\cite{greshake_indirect,injecagent}. Benchmarks such as InjecAgent and AgentDojo further show that such attacks are effective in realistic tool-integrated and multi-step agent settings~\cite{injecagent,agentdojo}.

Third, attacks may exploit feedback from tool execution. Tool-return injection manipulates the outputs returned by external tools and reinserts attacker-controlled content into the context, thereby influencing subsequent reasoning and action selection~\cite{agentdojo}. This attack path is especially relevant to autonomous agents because tool outputs are repeatedly consumed as intermediate feedback during long-horizon tasks.

Finally, recent studies show that context-layer attacks are becoming more adaptive and interaction-aware. ChatInject exploits structured chat templates and multi-turn interactions to improve attack success rates, while PISmith uses reinforcement learning to automatically discover stronger prompt injection strategies~\cite{chatinject,pismith}. These works indicate that the context and instruction layer is not only vulnerable to static malicious strings, but also to adaptive attacks that exploit the evolving structure of agent-context interaction.

\textbf{Defenses.}
Existing defenses primarily operate at the prompt or parsing level. Techniques such as instruction filtering, prompt delimiting, and structured prompting aim to separate trusted instructions from untrusted content~\cite{prompt_injection_survey,promptarmor,datafilter}. Trust-aware parsing and constrained query mechanisms further attempt to limit how external data is incorporated into the context. More recent approaches propose modular defenses such as input/output firewalls at the agent--tool interface to mitigate indirect injection risks~\cite{firewall_defense}.

\begin{table}[t]
\centering
\caption{Summary of attack surfaces, defenses, and limitations in the context and instruction layer.}
\label{tab:context_layer_mapping}
\tiny
\setlength{\tabcolsep}{1.3pt}
\renewcommand{\arraystretch}{1.15}
\begin{tabularx}{\columnwidth}{@{}p{0.19\columnwidth} X X X@{}}
\toprule
\textbf{Attack Surface} & \textbf{Representative Attacks} & \textbf{Corresponding Defenses} & \textbf{Limitations} \\
\midrule
Direct instruction channel 
& Direct prompt injection through malicious user inputs 
& Instruction filtering; prompt delimiting; structured prompting 
& Easily bypassed by adaptive or obfuscated instructions \\

External or retrieved content 
& Indirect prompt injection through web pages, documents, emails, or retrieved data 
& Trust-aware parsing; constrained retrieval; external-content filtering 
& Difficult to reliably distinguish untrusted data from task-relevant evidence \\

Tool feedback 
& Tool-return injection through malicious or manipulated tool outputs 
& Input/output firewalls; tool-output sanitization; context separation 
& Tool outputs are often repeatedly reused as intermediate feedback in multi-step tasks \\

Adaptive interaction 
& Multi-turn and template-aware injection, such as ChatInject and PISmith-style adaptive attacks 
& Robust prompt policies; adversarial testing; layered runtime monitoring 
& Prompt-level defenses remain brittle and may not generalize to adaptive adversaries \\
\bottomrule
\end{tabularx}
\end{table}

\textbf{Discussion.}
Overall, the context and instruction layer serves as the primary entry point for adversarial influence in autonomous agent frameworks. As summarized in Table~\ref{tab:context_layer_mapping}, existing defenses mainly attempt to separate trusted instructions from untrusted data through filtering, delimiting, parsing, and runtime monitoring. However, these mechanisms largely provide soft constraints during context construction and remain limited against adaptive adversaries and cross-layer propagation. In addition, many current studies still evaluate such risks in relatively short-horizon settings and do not fully capture how manipulated context propagates into tool execution, persistent state, or ecosystem-level consequences. A fundamental challenge is the lack of a formally enforced trust hierarchy within the context construction process, leaving agents inherently vulnerable to instruction-data ambiguity. This suggests that securing the context and instruction layer requires not only prompt-level safeguards, but also stronger coordination with protections for execution, state, and system-level control.

\subsection{Tool and Action Layer}
\label{sec:tool_action_layer}

\textbf{Layer Role.}
The tool and action layer defines the execution capability of the agent, enabling it to interact with external systems and perform real-world operations beyond language generation. It serves as the interface through which high-level decisions are realized as concrete actions, bridging the gap between abstract reasoning and environment interaction~\cite{yao2023react,schick2023toolformer,gorilla}.

\textbf{Attack Surface.}
Security risks in this layer arise from the coupling between natural language reasoning and structured tool invocation. Agents must select appropriate tools, generate valid arguments, and coordinate multi-step execution under an uncertain and potentially adversarial context. This introduces multiple attack surfaces, including tool selection, parameter construction, execution chaining, and resource consumption. Because tool descriptions and invocation interfaces are often exposed as natural language specifications, they can be manipulated similarly to prompt injection, but with direct consequences on system actions.

\textbf{Attacks.}
Attacks on the tool and action layer arise from different stages of tool-mediated execution.

First, tool selection attacks manipulate how agents choose among available tools. Since tool descriptions and usage instructions are often written in natural language, adversaries can modify or inject tool documentation to bias the agent toward an unsafe or attacker-controlled tool. For example, ToolHijacker shows that malicious tool documentation can systematically steer agents toward attacker-controlled tools~\cite{toolhijacker}. Empirical studies further show that LLM agents may exhibit unstable tool-selection preferences, where small perturbations in tool descriptions can alter the selected tool and expose exploitable decision boundaries~\cite{tool_preference}.

Second, parameter manipulation attacks target the argument construction process after a tool has been selected. Even when the chosen tool is legitimate, adversarial context can cause the agent to generate unsafe arguments, such as attacker-controlled URLs, file paths, commands, recipients, or queries. In tool-integrated agents, indirect prompt injection can propagate from external content into tool arguments, leading to unintended operations, privacy leakage, or data exfiltration~\cite{injecagent}.

Third, execution-chain attacks exploit the multi-step nature of agent workflows. Rather than compromising a single tool call, adversaries can steer intermediate reasoning steps or tool outputs so that later invocations gradually deviate from the user's original intent. This attack pattern is particularly relevant to autonomous agents because tool calls, observations, and follow-up actions are repeatedly interleaved during long-horizon execution.

Finally, resource-amplification attacks exploit tool-calling loops and execution chains to increase operational cost. For example, Clawdrain demonstrates that adversaries can manipulate tool-use workflows to amplify token consumption, achieving 6--7$\times$ token usage and up to 9$\times$ in certain configurations~\cite{clawdrain}. These attacks show that tool-layer vulnerabilities affect not only task correctness and safety, but also the economic and operational reliability of agent systems.

\textbf{Defenses.}
Defenses at this layer focus on constraining execution and reducing the attack surface of tool interaction. Common approaches include tool allowlisting, argument validation, and execution monitoring, which aim to restrict unsafe actions or detect anomalous behavior. Sandboxing mechanisms further isolate tool execution environments to limit the impact of malicious operations. Recent work also explores structured tool invocation protocols and function masking to reduce ambiguity in tool usage~\cite{hammer}. More recently, TrustAgent proposes an Agent-Constitution framework that enforces pre-planning and post-action safety checks to constrain agent behavior within predefined safety boundaries~\cite{trustagent}, while ToolEmu uses LM-emulated sandboxes to proactively identify tool-use risks before real-world deployment~\cite{toolemu}. R-Judge further contributes a benchmark for evaluating LLM agents' safety risk awareness during tool-mediated execution~\cite{rjudge}.

\begin{table}[t]
\centering
\caption{Summary of attack surfaces, defenses, and limitations in the tool and action layer.}
\label{tab:tool_layer_mapping}
\tiny
\setlength{\tabcolsep}{1.3pt}
\renewcommand{\arraystretch}{1.15}
\begin{tabularx}{\columnwidth}{@{}p{0.19\columnwidth} X X X@{}}
\toprule
\textbf{Attack Surface} & \textbf{Representative Attacks} & \textbf{Corresponding Defenses} & \textbf{Limitations} \\
\midrule
Tool selection 
& ToolHijacker-style manipulation of tool descriptions; unstable tool preferences 
& Tool allowlisting; trusted tool registries; tool metadata verification 
& Restricts flexibility and is hard to maintain in dynamic tool ecosystems \\

Parameter construction 
& Malicious arguments such as attacker-controlled URLs, file paths, commands, recipients, or queries 
& Argument validation; schema checking; policy-based constraints 
& Difficult to generalize across heterogeneous tools and task-specific arguments \\

Execution chaining 
& Multi-step deviation through intermediate reasoning, tool outputs, or action chains 
& Execution monitoring; step-wise approval; structured tool invocation 
& Often reactive and may miss delayed or gradual deviations across long-horizon workflows \\

Resource consumption 
& Clawdrain-style token or tool-call amplification through repeated tool-use workflows 
& Rate limiting; budget control; sandboxing 
& May reduce useful long-horizon execution and cannot fully distinguish malicious from expensive benign tasks \\
\bottomrule
\end{tabularx}
\end{table}

\textbf{Discussion.}
Overall, attacks in this layer show how adversarial influence can move from language-level manipulation to concrete executable actions. As summarized in Table~\ref{tab:tool_layer_mapping}, existing defenses mainly attempt to constrain tool availability, validate arguments, monitor execution behavior, and isolate execution environments. However, these mechanisms are often more effective at limiting immediate action risks than at preventing upstream manipulation of tool selection or downstream propagation through long-horizon workflows. A central challenge is that many defenses intervene only after an action is generated or executed, whereas attacks may begin earlier during tool selection, parameter construction, or intermediate reasoning. In addition, current studies still focus largely on isolated tool calls or relatively short execution chains, while less attention has been paid to how tool-level manipulation interacts with persistent state, automation mechanisms, and ecosystem-level trust boundaries. This suggests that securing the tool and action layer requires not only better execution constraints, but also stronger decision-time safeguards and tighter coordination with protections for state and ecosystem-level components.
\subsection{State and Persistence Layer}
\label{sec:state_persistence_layer}

\textbf{Layer Role.}
The state and persistence layer governs how an agent retains and reuses information over time, enabling continuity across multiple interaction steps. It provides the temporal foundation for long-horizon reasoning, allowing past actions and observations to influence future decisions. This persistent state differentiates autonomous agents from stateless LLM applications and plays a central role in shaping their evolving behavior~\cite{memgpt,park2023generative}.

\textbf{Attack Surface.}
Security risks in this layer arise from the persistence and reuse of intermediate information. Agent systems often store interaction histories, retrieved knowledge, and tool outputs, which are later reintroduced into the context. If malicious or manipulated content is stored, it can influence future decisions over extended time horizons. This creates an attack surface where adversaries can inject harmful information once and achieve persistent effects across multiple interactions, leading to compounding behavioral deviations.

\textbf{Attacks.}
Attacks on the state and persistence layer arise from how adversarial information is stored, accumulated, reused, and carried across sessions.

First, memory poisoning attacks inject malicious or misleading content into the agent's persistent memory store. Once written into memory, such content may be retrieved in later reasoning steps and treated as relevant background information, thereby influencing subsequent decisions and actions~\cite{mem_poisoning}. Unlike prompt injection, which typically affects a single interaction, memory poisoning can persist beyond the original attack session.

Second, state accumulation attacks exploit the way agents continuously collect and summarize interaction histories, retrieved knowledge, observations, and reflections. In generative agent frameworks, for example, stored observations and reflections shape long-term behavior, which means compromised state can gradually bias future planning and decision-making~\cite{park2023generative}. This risk becomes more severe when agents repeatedly reuse historical context without reliable provenance tracking or validation.

Third, memory-mediated indirect injection attacks occur when adversarial content from external sources is first stored as state and later reintroduced into the context. In long-context or tool-integrated agents, indirect prompt injection can therefore propagate through memory, turning transient untrusted content into a persistent influence channel~\cite{injecagent}.

Finally, cross-session attacks exploit weak isolation between interactions or tasks. Adversarial inputs introduced in earlier sessions may affect later sessions if the agent retrieves poisoned memories or stale workspace state, making it difficult to contain attacks within a single execution boundary~\cite{memory_attack_survey}.

\textbf{Defenses.}
Defenses for this layer focus on controlling how information is stored, retrieved, and reused. Common approaches include memory filtering, relevance-based retrieval, and decay mechanisms that reduce the influence of older or low-confidence information. Some systems incorporate memory auditing or provenance tracking to identify potentially malicious entries~\cite{mempoison}. In the context of retrieval-augmented generation (RAG), TrustRAG proposes a defense framework that detects and filters poisoned retrieval content to enhance the robustness and trustworthiness of knowledge-grounded agent behavior~\cite{trustrag}.

\begin{table}[t]
\centering
\caption{Summary of attack surfaces, defenses, and limitations in the state and persistence layer.}
\label{tab:state_layer_mapping}
\tiny
\setlength{\tabcolsep}{1.3pt}
\renewcommand{\arraystretch}{1.15}
\begin{tabularx}{\columnwidth}{@{}p{0.19\columnwidth} X X X@{}}
\toprule
\textbf{Attack Surface} & \textbf{Representative Attacks} & \textbf{Corresponding Defenses} & \textbf{Limitations} \\
\midrule
Memory storage 
& Memory poisoning through malicious or misleading persistent entries 
& Memory filtering; write-time validation; memory sanitization 
& Subtle or context-dependent poisoned content is difficult to detect \\

State accumulation 
& Gradual contamination of interaction histories, retrieved knowledge, observations, or reflections 
& Relevance-based retrieval; confidence scoring; periodic memory review 
& Poisoned state may still appear relevant and influence long-term reasoning \\

Memory reuse 
& Memory-mediated indirect injection where external content is stored and later reintroduced into context 
& Provenance tracking; source-aware retrieval; context separation 
& Provenance is costly to maintain and may not fully prevent unsafe reuse \\

Cross-session persistence 
& Attacks that survive across sessions through poisoned memories or stale workspace state 
& Memory decay; session isolation; memory auditing 
& Strong isolation or decay may remove useful long-term context and reduce agent capability \\
\bottomrule
\end{tabularx}
\end{table}

\textbf{Discussion.}
Overall, attacks in this layer show that adversarial influence can persist, accumulate, and reappear across multiple reasoning steps or sessions. As summarized in Table~\ref{tab:state_layer_mapping}, existing defenses mainly attempt to filter stored content, constrain memory retrieval, reduce stale influence, or audit memory provenance. However, these mechanisms mostly mitigate immediate or local memory risks and provide limited guarantees for long-horizon state integrity. A central challenge is the tension between memory utility and memory security: defenses that reduce persistent influence may also weaken the long-term context that autonomous agents rely on. In addition, current studies still provide limited analysis of delayed activation, recovery after compromise, and the interaction between poisoned state and downstream tool execution. This suggests that securing the state and persistence layer requires not only better memory sanitization and auditing, but also stronger mechanisms for long-term state isolation, provenance reasoning, and recovery across extended agent lifecycles.

\subsection{Ecosystem and Automation Layer}
\label{sec:ecosystem_automation_layer}

\textbf{Layer Role.}
The ecosystem and automation layer governs how an autonomous agent extends its capabilities beyond the core system and interacts with external environments. It enables integration with third-party services, tools, and platforms, allowing the agent to access resources and functionalities that are not natively available within the model itself. This layer also supports automated operation through mechanisms such as scheduled tasks, event-driven execution, and coordination across multiple agents, enabling continuous and large-scale deployment~\cite{gorilla,thirdparty_api_attack}.

\textbf{Attack Surface.}
Security risks in this layer arise from the openness and composability of the agent ecosystem. Agents often rely on external components whose behavior may not be fully controlled or verified. This includes third-party skills, plugin marketplaces, and external APIs, which may introduce vulnerabilities through malicious functionality or unintended side effects. Furthermore, automation mechanisms enable actions to be executed without immediate human oversight, allowing attacks to propagate over time or across interconnected systems. These characteristics create an attack surface that extends beyond individual agents to the broader ecosystem.

\textbf{Attacks.}
Attacks on the ecosystem and automation layer arise from how adversaries exploit external extensions, trust relationships, automated execution, and cross-system propagation.

First, supply-chain attacks target third-party skills, plugins, or marketplace components. In this setting, adversaries introduce malicious functionality into seemingly benign extensions, which may later be invoked by agents as trusted capabilities~\cite{plugin_attack}. These attacks exploit the openness of agent ecosystems and the difficulty of verifying external components before integration.

Second, trust-boundary attacks exploit weak isolation among agents, tools, APIs, and shared resources. When agents compose multiple ecosystem components, they may implicitly trust outputs, permissions, or assumptions from one component and reuse them in another. This creates propagation paths through which a compromised plugin, API response, or shared artifact can influence broader agent behavior.

Third, automation-based attacks exploit scheduled, event-driven, or unattended execution mechanisms. Adversaries may inject malicious triggers, manipulate execution conditions, or poison recurring workflows so that agents repeatedly execute harmful actions without immediate human oversight. Compared with one-shot attacks, automation abuse can amplify impact over time and make failures harder to detect.

Fourth, multi-agent attacks exploit communication and coordination among agents. In multi-agent systems, a compromised agent can influence others through shared messages, delegated tasks, or collaborative planning channels, allowing adversarial behavior to spread beyond a single agent boundary~\cite{redteam_multiagent}.

Finally, platform-scale API attacks arise from large tool and API ecosystems. As API-integrated models connect to many external services, improper handling of API behavior, permissions, and dependencies can lead to cascading failures or unintended cross-system interactions~\cite{gorilla}. These findings show that ecosystem-layer vulnerabilities are not confined to individual agents, but may propagate across connected components and affect entire agent platforms.

\textbf{Defenses.}
Defenses for this layer focus on establishing trust and control over ecosystem components and automated execution. Common approaches include permission control, sandboxing of plugins and services, and capability-based access restriction. Verification mechanisms such as code auditing and reputation systems are also used to assess the trustworthiness of third-party components. For automation, safeguards such as rate limiting, execution monitoring, and human-in-the-loop verification can reduce the risk of uncontrolled actions. Recent work has begun to address these challenges more systematically. AGrail proposes a lifelong agent guardrail with adaptive safety detection that can operate across different layers and evolving tasks~\cite{agrail}. Dong et al.\ provide a comprehensive survey of guardrails and safeguard mechanisms for large language models, covering both input-level and system-level protections~\cite{safeguarding_llms}. Zhao et al.\ further analyze security risks arising from third-party API integrations in LLM ecosystems, highlighting plugin-level attack vectors and potential countermeasures~\cite{thirdparty_api_attack}.

\begin{table}[t]
\centering
\caption{Summary of attack surfaces, defenses, and limitations in the ecosystem and automation layer.}
\label{tab:ecosystem_layer_mapping}
\tiny
\setlength{\tabcolsep}{1.3pt}
\renewcommand{\arraystretch}{1.15}
\begin{tabularx}{\columnwidth}{@{}p{0.19\columnwidth} X X X@{}}
\toprule
\textbf{Attack Surface} & \textbf{Representative Attacks} & \textbf{Corresponding Defenses} & \textbf{Remaining Limitations} \\
\midrule
External extensions 
& Supply-chain attacks through malicious skills, plugins, or marketplace components 
& Plugin verification; code auditing; reputation systems 
& Difficult to scale in dynamic ecosystems with many third-party components \\

Trust boundaries 
& Compromised plugins, API responses, or shared artifacts influencing other components 
& Permission control; capability-based access restriction; isolation policies 
& Coarse-grained permissions may not capture complex cross-component dependencies \\

Automated execution 
& Scheduled, event-driven, or recurring workflows repeatedly executing harmful actions 
& Rate limiting; execution monitoring; human-in-the-loop verification 
& Often reactive and may not prevent delayed or unattended abuse \\

Multi-agent coordination 
& Compromised agents influencing others through messages, delegated tasks, or collaborative planning 
& Agent authentication; communication monitoring; coordination policies 
& Trust management across agents remains under-specified and hard to enforce \\

Platform-scale APIs 
& Cascading failures or unintended interactions across large API ecosystems 
& API sandboxing; dependency tracking; secure integration protocols 
& Hard to reason about systemic effects across heterogeneous services \\
\bottomrule
\end{tabularx}
\end{table}

\textbf{Discussion.}
Overall, attacks in this layer show that security risks can arise not only from individual agent behavior, but also from the broader ecosystem in which agents are deployed. As summarized in Table~\ref{tab:ecosystem_layer_mapping}, existing defenses mainly attempt to establish trust and control through permission restriction, sandboxing, component verification, execution monitoring, and human oversight. However, these mechanisms mainly reduce local or component-level risks and remain limited in governing dynamic, composable, and large-scale agent ecosystems. A central challenge is that defenses often target individual components, whereas ecosystem-layer attacks exploit relationships among components, agents, services, and automated workflows. In addition, existing studies remain relatively sparse and often lack standardized threat models for plugin supply chains, ecosystem trust boundaries, automated execution, and cross-agent propagation. This suggests that securing the ecosystem and automation layer requires stronger trust management, more explicit integration boundaries, and better system-level governance mechanisms for external extensions and automated workflows.
\section{Cross-Layer Attacks and Defenses}
\label{sec:Cross-Layer Attacks and Defenses}
While Section~\ref{sec:Layered Attacks and Defenses} reviews security risks and defense mechanisms within each layer, this section provides a cross-layer synthesis of the existing literature. Specifically, we organize the discussion around three questions: how representative studies are distributed across the proposed layers, how attacks propagate across layer boundaries, and how current defenses align with or fail to address such cross-layer threats. First, we compare primary studies across the context, tool, state, and ecosystem layers to reveal the current research landscape. Second, we analyze an end-to-end attack propagation process that connects input manipulation, action execution, state persistence, and ecosystem-level amplification. Third, we compare defense strategies across layers to identify gaps between existing safeguards and the evolving threat model. This synthesis motivates the open problems and future directions discussed in Section~\ref{sec:Open Problems and Future Directions}.
\subsection{Representative Studies Across Layers}

To provide a unified view of existing research, we summarize representative studies across different layers of autonomous agent frameworks. 
Rather than treating each layer in isolation, this comparison highlights how prior work is distributed across context, tool, state, and ecosystem layers, and reveals the current focus and limitations of the research landscape.
\begin{table*}[!htbp]
\centering
\caption{Representative studies across different layers of autonomous agent frameworks.}
\label{tab:layer_comparison}
\footnotesize
\setlength{\tabcolsep}{3pt}
\renewcommand{\arraystretch}{1.15}
\begin{threeparttable}
\begin{tabular}{l c l l c c p{3.6cm}}
\toprule
\textbf{Study} & \textbf{Year} & \textbf{Layer} & \textbf{Attack Focus} & \textbf{Def.} & \textbf{Eval.} & \textbf{Key Idea} \\
\midrule
\multicolumn{7}{l}{\textbf{Context and Instruction Layer}} \\
Prompt Injection~\cite{prompt_injection} & 2023 & Context & Direct injection & N & Emp. & Override system instructions via crafted prompts \\
Indirect Injection~\cite{greshake_indirect} & 2023 & Context & Indirect injection & N & Emp. & Embed malicious instructions in external data \\
Adaptive Attacks~\cite{adaptive_attack_ipi} & 2025 & Context & Adaptive attack & N & Emp. & Existing IPI defenses can be bypassed \\
InjecAgent~\cite{injecagent} & 2024 & Context / Tool & Indirect injection & N & Bench. & Injection propagation via tool usage \\
AgentDojo~\cite{agentdojo} & 2024 & Context / Tool & Benchmark & P & Bench. & Evaluate injection in realistic agent tasks \\
Trojan's Whisper~\cite{liu2026trojanswhisper} & 2026 & Context / State & Bootstrap injection & N & Emp. & Persistent poisoning via startup artifacts \\
TrustRAG~\cite{trustrag} & 2025 & Context / State & Defense & Y & Emp. & Filter poisoned retrievals for robust RAG \\
\midrule
\multicolumn{7}{l}{\textbf{Tool and Action Layer}} \\
ToolHijacker~\cite{toolhijacker} & 2025 & Tool & Tool selection & N & Emp. & Manipulate tool choice via descriptions \\
ToolSword~\cite{toolsword} & 2024 & Tool & Safety evaluation & N & Bench. & Three-stage tool safety analysis \\
ToolEmu~\cite{toolemu} & 2023 & Tool & Risk evaluation & N & Bench. & LM-emulated sandbox for risk identification \\
TrustAgent~\cite{trustagent} & 2024 & Tool & Defense & Y & Bench. & Safety-aware planning with action guardrails \\
Clawdrain~\cite{clawdrain} & 2026 & Tool / Eco. & Resource abuse & N & Emp. & Amplify tool calls to exhaust resources \\
Third-party API~\cite{thirdparty_api_attack} & 2024 & Tool / Eco. & API manipulation & N & Emp. & Exploit third-party APIs in LLM pipelines \\
\midrule
\multicolumn{7}{l}{\textbf{State and Persistence Layer}} \\
MemPoison~\cite{mempoison} & 2026 & State & Memory poisoning & N & Emp. & Inject persistent adversarial memory \\
Generative Agents~\cite{park2023generative} & 2023 & State & Long-horizon effect & N & Emp. & Behavior shaped by stored memory \\
\midrule
\multicolumn{7}{l}{\textbf{Ecosystem and Automation Layer}} \\
Plugin Attacks~\cite{plugin_attack} & 2026 & Ecosystem & Supply-chain & N & Theo. & Supply-chain poisoning in agent skill ecosystems \\
Red-Teaming MAS~\cite{redteam_multiagent} & 2025 & Ecosystem & Communication attack & N & Emp. & Attacks via inter-agent messages \\
\midrule
\multicolumn{7}{l}{\textbf{Cross-Layer}} \\
ASB~\cite{zhang2025asb} & 2025 & Cross-layer & Benchmark & P & Bench. & Unified formalization of agent attacks and defenses \\
PASB~\cite{wang2026pasb} & 2026 & Cross-layer & Benchmark & N & Bench. & Formalized attacks on OpenClaw agents \\
\bottomrule
\end{tabular}
\begin{tablenotes}
\scriptsize
\item Def. = Defense Proposed (Y = Yes, P = Partial, N = No); Eval. = Evaluation Method (Emp. = Empirical, Bench. = Benchmark, Theo. = Theoretical); Eco. = Ecosystem; MAS = Multi-Agent Systems.
\end{tablenotes}
\end{threeparttable}
\end{table*}

The comparison reveals several notable trends. As shown in Table~\ref{tab:layer_comparison}, existing research remains heavily concentrated on the context and tool layers, which together account for 13 out of 19 representative studies, while the state layer (2 studies) and ecosystem layer (2 studies) receive comparatively less attention. Second, the defense landscape is uneven: only 2 out of 19 studies propose full defense mechanisms (TrustRAG and TrustAgent), while 2 others offer partial defenses (AgentDojo and ASB), and most existing defenses target the context or tool layers. Third, recent studies increasingly consider cross-layer interactions, such as the propagation of indirect injection through tool usage~\cite{injecagent,agentdojo} and bootstrap-based persistent attacks~\cite{liu2026trojanswhisper}. Finally, system-level risks, including persistent memory manipulation and ecosystem vulnerabilities, remain under-explored, highlighting important directions for future research.
\subsection{Cross-Layer Attacks}
\label{sec:cross_layer_attack_propagation}
\begin{figure}[t]
    \centering
    \includegraphics[width=\columnwidth]{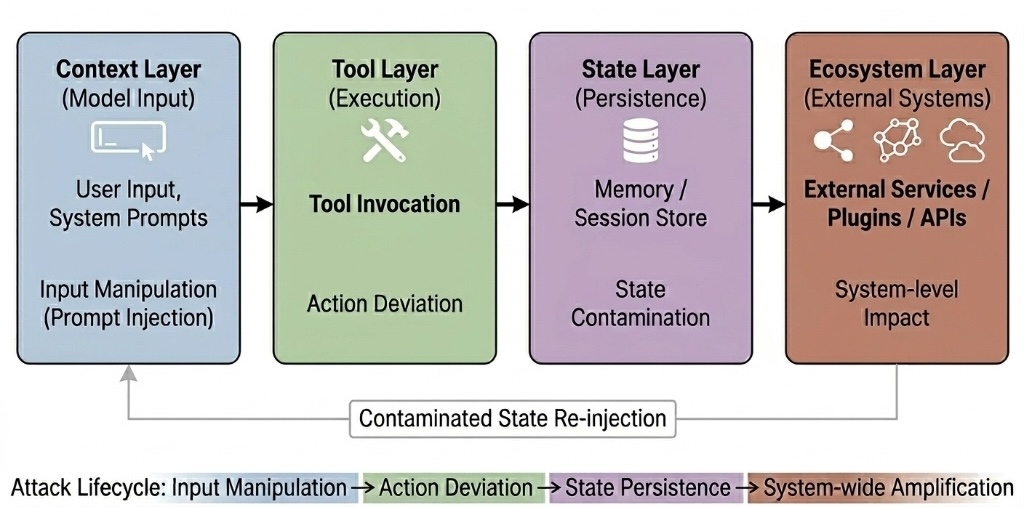}
    \caption{Cross-layer attack propagation in autonomous agent frameworks.}
    \label{fig:cross_layer_attack}
\end{figure}
As illustrated in Figure~\ref{fig:cross_layer_attack}, attacks in autonomous agent frameworks may propagate across layers rather than remaining confined to a single component. The propagation typically begins at the context and instruction layer, where adversarial content is introduced into the agent's input through direct prompt injection, indirect injection via external sources, or manipulated tool returns. Because the agent's reasoning relies on the composite context assembled at each step, such injected content can alter the agent's interpretation of task goals, override safety constraints, or introduce fabricated instructions, ultimately producing manipulated decisions that deviate from the user's original intent~\cite{prompt_injection,greshake_indirect,agentdojo}. These manipulated decisions then directly affect the tool and action layer. Once the agent selects tools or constructs action chains based on compromised reasoning, the resulting behavior may deviate from intended objectives. Recent studies show that indirect prompt injection can propagate through tool usage, leading to biased tool selection or unsafe API invocation~\cite{injecagent,agentdojo}. In this stage, attacks transition from influencing model outputs to controlling executable actions, marking a critical escalation from reasoning-level manipulation to real-world effects.

The consequences of these actions are then recorded in the state and persistence layer. Outputs generated by compromised executions may be stored in memory, session traces, or workspace states, where they can persist across multiple interactions. This enables attacks such as memory poisoning and state contamination, where adversarial information becomes part of the agent's long-term context~\cite{mempoison,park2023generative}. For example, Trojan's Whisper demonstrates that adversarial instructions embedded in startup artifacts can persistently poison the agent's context across sessions~\cite{liu2026trojanswhisper}. Unlike transient prompt-level attacks, these persistent effects can influence future behavior even after the original attack vector is removed.

Finally, the impact may extend to the ecosystem and automation layer. Through integrations with external services, plugins, and automation mechanisms, compromised states and actions can propagate beyond a single agent instance. Supply-chain risks in plugin ecosystems, automation abuse via scheduled tasks, and multi-agent coordination further amplify the attack impact across systems, as recent red-teaming studies confirm that attacks can propagate through inter-agent communication channels~\cite{thirdparty_api_attack,redteam_multiagent}. At this stage, vulnerabilities evolve into system-level risks, affecting broader environments and potentially multiple interconnected agents.

Notice this propagation is not strictly linear. Feedback loops may exist in which contaminated state information is reintroduced into the context layer, forming iterative attack cycles. Such cycles enable adversaries to reinforce and sustain their influence over time, making detection and mitigation more challenging.

Overall, the security of autonomous agent frameworks cannot be understood solely at the level of individual components. Instead, it requires a cross-layer perspective that captures how attacks originate, propagate, persist, and amplify. This layered propagation mechanism highlights the need for coordinated defenses that operate across context construction, action execution, state management, and ecosystem governance.
\subsection{Cross-Layer Defenses}

Although a wide range of defense mechanisms have been proposed for autonomous agent frameworks, a cross-layer comparison reveals several fundamental mismatches between existing defenses and the evolving attack landscape.

First, there exists a clear imbalance in defense strength across layers. Most existing work focuses on prompt-level safeguards, such as input filtering and structured prompting, which provide only soft constraints on model behavior~\cite{prompt_injection_survey}, and recent work shows that such defenses can be bypassed by adaptive attack strategies~\cite{adaptive_attack_ipi}. In contrast, stronger enforcement mechanisms, such as capability restriction and sandboxing, are typically applied at the tool or system level, as exemplified by TrustAgent's safety-aware planning with action guardrails~\cite{trustagent}. At the state layer, TrustRAG addresses retrieval poisoning through filtering mechanisms~\cite{trustrag}. However, defenses targeting persistent state and long-horizon behavior remain relatively underexplored overall, leaving a critical gap in protecting against memory-based attacks.

Second, current defenses are largely designed in isolation and do not account for cross-layer attack propagation. While individual mechanisms may be effective within a single layer, they often fail to prevent attacks that originate in one layer and manifest in another. For example, prompt-level filtering may reduce injection attempts, but cannot prevent manipulated inputs from influencing tool selection and execution ~\cite{greshake_indirect, injecagent}. This mismatch highlights the limitation of layer-specific defenses in the presence of tightly coupled reasoning-action-memory loops.

Third, existing approaches struggle to address persistent adversarial effects. Once malicious information is stored in the agent's state, it can influence future behavior independently of the original attack vector. Current defenses provide limited guarantees on state integrity and lack standardized mechanisms for tracking or validating long-term memory ~\cite{mempoison, park2023generative}. As a result, attacks can persist and evolve over time, even in the presence of input-level protections.

Finally, deployment-oriented challenges further limit the effectiveness of existing defenses. Techniques such as sandboxing, permission control, and execution monitoring require careful configuration and often introduce trade-offs between security and system usability ~\cite{thirdparty_api_attack}. In open and extensible ecosystems, enforcing consistent security policies across third-party components remains particularly difficult.

Overall, these observations suggest that current defense strategies are not aligned with the cross-layer nature of attacks in autonomous agent frameworks. Addressing this gap requires moving beyond isolated safeguards toward more holistic and coordinated defense mechanisms, which we discuss in Section~\ref{sec:Open Problems and Future Directions}.

\subsection{Case Study: A Plausible Cross-Layer Attack Scenario in OpenClaw}

To illustrate how the layered taxonomy can be applied to a real autonomous agent framework, we present a plausible cross-layer attack scenario grounded in publicly documented OpenClaw mechanisms. Unlike the abstract propagation pattern discussed in Section~\ref{sec:cross_layer_attack_propagation}, this case study focuses on how a concrete attack chain could unfold through OpenClaw's runtime prompt construction, typed tool interfaces, persistent workspace state, and automation mechanisms~\cite{openclaw_systemprompt,openclaw_context,openclaw_tools,openclaw_workspace,openclaw_heartbeat,openclaw_cron}.

In this scenario, the attack starts at the context and instruction layer. OpenClaw assembles model-visible context from multiple runtime components, including system prompts, tool descriptions, skill information, session context, and workspace artifacts~\cite{openclaw_systemprompt,openclaw_context}. If adversarial instructions are embedded in a workspace file such as \texttt{CLAUDE.md}, a bootstrap artifact, or another file later referenced during execution, they may be incorporated into the composite context and interpreted by the model as actionable guidance. This risk is consistent with the broader literature on indirect prompt injection, in which untrusted external content is transformed into an implicit instruction source once merged into the model's input~\cite{greshake_indirect,injecagent}.

The manipulated context can then steer the tool and action layer. OpenClaw exposes executable capabilities through typed tools rather than plain text outputs~\cite{openclaw_tools}. Under poisoned contextual guidance, the agent may invoke a Bash tool, file-writing tool, or another execution interface using attacker-influenced arguments. In this way, the attack moves from prompt-level manipulation to concrete system actions, similar to how indirect injection can propagate into unsafe tool invocation in tool-integrated agents~\cite{injecagent,agentdojo}. The key issue here is not only whether a tool is available, but whether the selected tool and its parameters remain aligned with the user's original intent.

The resulting effects can persist through the state and persistence layer. OpenClaw maintains agent-specific workspaces and related runtime state that may survive across interactions~\cite{openclaw_workspace}. If compromised tool execution writes attacker-influenced content into workspace files or other persistent artifacts, that content can later be reloaded into future sessions and once again become part of the agent's effective context. This creates a persistence channel analogous to memory poisoning and long-horizon state contamination, where adversarial influence survives beyond the initial triggering interaction and continues shaping later behavior~\cite{mempoison,park2023generative,memory_attack_survey}.

The attack may then be amplified at the ecosystem and automation layer. OpenClaw supports recurring execution through mechanisms such as heartbeat and cron~\cite{openclaw_heartbeat,openclaw_cron}. If a compromised workflow or poisoned artifact becomes part of an automated routine, the same attacker-influenced behavior may be repeatedly re-executed without immediate human review. Moreover, if the workflow interacts with external services, plugins, or connected tools, the impact can extend beyond a single agent run and become an ecosystem-level risk. This progression mirrors broader concerns about automation abuse, supply-chain exposure, and multi-agent or multi-service propagation in agent ecosystems~\cite{thirdparty_api_attack,redteam_multiagent}.

This scenario also clarifies the practical meaning of defense gaps in OpenClaw-like systems. At the context layer, runtime permission prompts and execution controls do not by themselves resolve the trust ambiguity of workspace artifacts once they are incorporated into model-visible context~\cite{openclaw_systemprompt,openclaw_context}. At the tool layer, allowlisting and sandboxing can restrict available operations, but they do not necessarily guarantee that tool arguments are semantically safe or aligned with task intent~\cite{openclaw_tools,hammer}. At the state layer, persistent workspace artifacts can continue carrying adversarial influence in the absence of stronger provenance tracking, sanitization, or recovery mechanisms~\cite{openclaw_workspace,mempoison}. At the automation layer, heartbeat and cron may include scheduling or rate-control mechanisms, yet these controls are not equivalent to content-aware monitoring of whether a recurring workflow has already been compromised~\cite{openclaw_heartbeat,openclaw_cron}.

Overall, this OpenClaw-based walkthrough provides a concrete instantiation of the cross-layer propagation pattern illustrated in Figure~\ref{fig:cross_layer_attack}. More importantly, it shows why the layered taxonomy is analytically useful: the main security issue is not a single isolated vulnerability, but the way context construction, tool invocation, persistent state, and automation mechanisms can combine into an end-to-end attack chain. In this sense, OpenClaw is not merely an architectural example, but a structured case through which the practical value of the proposed layered analysis can be demonstrated.

\section{Open Problems and Future Directions}
\label{sec:Open Problems and Future Directions}
Building on the layered analysis in Section~\ref{sec:Layered Attacks and Defenses} and the cross-layer comparison in Section~\ref{sec:Cross-Layer Attacks and Defenses}, this section discusses the major open problems and potential future directions for security research on autonomous agent frameworks. We focus on three key challenges: research imbalance across layers, lack of long-horizon evaluation, and weak ecosystem trust models. Each subsection corresponds to one open problem: we first describe the problem and the challenges it creates for current research, and then discuss possible directions for addressing it.

\subsection{Research Imbalance Across Layers}

A fundamental open problem in autonomous agent security is the uneven distribution of research attention across different architectural layers. As highlighted in Section~\ref{sec:Cross-Layer Attacks and Defenses}, most existing studies focus heavily on the context and instruction layer, particularly prompt injection attacks and their variants~\cite{prompt_injection_survey,greshake_indirect}. This focus is partly because context-level attacks are easier to model, evaluate, and reproduce, as they often involve manipulating model-visible inputs within a single interaction. However, real-world agent systems also rely on persistent state, long-horizon execution, external tools, automation mechanisms, and ecosystem integrations. Compared with prompt-level vulnerabilities, risks in the state and persistence layer and the ecosystem and automation layer remain less systematically studied, even though recent work has begun to examine memory poisoning, long-horizon manipulation, supply-chain attacks, and cross-agent propagation~\cite{mempoison,park2023generative,thirdparty_api_attack,redteam_multiagent}. This imbalance makes it difficult to obtain a complete understanding of agent security and may create a false sense of protection if defenses focus mainly on input-level safeguards.

Addressing this problem requires a more balanced research agenda across all layers of autonomous agent frameworks. Future work should develop benchmarks, threat models, and defense mechanisms for underexplored layers, especially state integrity, persistent memory, automation abuse, and ecosystem-level trust. For the state and persistence layer, research should investigate how adversarial information is stored, retrieved, forgotten, audited, and recovered over long time horizons. For the ecosystem and automation layer, future studies should examine plugin verification, capability control, cross-agent communication, and supply-chain risks in dynamic environments. More broadly, balanced layer-wise research should be combined with cross-layer analysis, so that defenses can account not only for isolated prompt-level attacks but also for how threats propagate through tools, state, and external ecosystems.

\subsection{Lack of Long-Horizon Evaluation}

A second open problem is the lack of long-horizon evaluation for autonomous agent security. Despite the growing body of research on agent attacks and defenses, most existing benchmarks remain limited to single-turn or short-horizon settings, focusing on immediate outcomes such as prompt injection success rates or unsafe tool invocation~\cite{agentdojo,injecagent}. While such metrics are useful for measuring local vulnerabilities, they do not adequately capture security risks that emerge over extended interaction horizons. As discussed in Sections~\ref{sec:state_persistence_layer} and~\ref{sec:cross_layer_attack_propagation}, attacks such as memory poisoning and state contamination may not produce immediate visible effects, but can gradually influence the agent's behavior through accumulated state~\cite{mempoison,park2023generative}. Similarly, cross-layer propagation can cause small initial manipulations to evolve into significant system-level deviations over time. This creates a major evaluation challenge: defenses that appear effective in short, isolated test cases may fail when adversarial influence persists, reappears, or compounds across multiple interactions.

Addressing this challenge requires evaluation frameworks that explicitly model long-term agent behavior. Future benchmarks should incorporate multi-step tasks, persistent memory, repeated interactions, and cross-session scenarios, enabling researchers to assess how attacks evolve and accumulate over time. Evaluation metrics should also go beyond immediate attack success rates to measure behavioral drift, persistence of adversarial effects, delayed impact, recovery capability after compromise, and resilience under continuous operation. In addition, long-horizon evaluation should consider how attacks move across context, tools, state, and ecosystem components, rather than treating each test case as an isolated interaction. Such evaluation frameworks are essential for understanding the true security risks of autonomous agent frameworks and for designing defenses that remain robust under realistic long-term deployment conditions.
\subsection{Weak Ecosystem Trust Model}

A third open problem is the weak ecosystem trust model in autonomous agent frameworks. Modern agent systems integrate external capabilities through APIs, skills, plugins, and third-party services, often sourced from heterogeneous and dynamically evolving environments~\cite{gorilla,thirdparty_api_attack}. While these integrations significantly enhance functionality, they also introduce supply-chain risks, where malicious or compromised components can execute unintended actions or leak sensitive information. This problem is further amplified by automation mechanisms such as scheduled execution, event-driven triggers, and multi-agent coordination, which allow agents to operate autonomously over extended periods. Once a malicious component is introduced or a trust assumption is violated, automated workflows may repeatedly execute harmful actions without immediate human oversight, causing small ecosystem-level vulnerabilities to escalate into large-scale failures~\cite{redteam_multiagent}. A key challenge is the absence of explicit and enforceable trust boundaries: current systems often rely on coarse-grained permission controls, sandboxing, or static trust assumptions, which provide limited guarantees in dynamic, composable, and interconnected environments.

Addressing this challenge requires rethinking trust management for agent ecosystems. Future research should explore fine-grained trust models, capability-based security mechanisms, and dynamic trust evaluation frameworks that continuously assess the behavior and reliability of external components. Standardized protocols for plugin verification, provenance tracking, permission specification, and secure integration are also needed to reduce risks introduced by third-party dependencies. In addition, ecosystem-level defenses should account for automation and multi-agent coordination by supporting runtime monitoring, rate limiting, human-in-the-loop approval for high-risk actions, and adaptive trust updates based on observed behavior. Such mechanisms can help move agent ecosystems from implicit and static trust assumptions toward explicit, enforceable, and continuously updated trust models.

\section{Conclusion}
\label{sec:conclusion}
In this survey, we presented a structured review of security risks in autonomous agent frameworks from a layered perspective, organizing existing work across four key layers: context and instruction, tool and action, state and persistence, and ecosystem and automation. Our analysis highlights that security risks are inherently cross-layer, where attacks can propagate from input manipulation to action execution, persist in system state, and amplify across interconnected environments. This reveals a fundamental mismatch between evolving threats and existing defenses, which are largely designed in isolation. We further identify key challenges, including research imbalance across layers, the lack of long-horizon evaluation, and weak ecosystem trust models. Addressing these issues calls for a shift toward integrated, system-level approaches, including cross-layer defense mechanisms, long-term evaluation frameworks, and trustworthy ecosystem design, to support the development of secure and reliable autonomous agent frameworks.

\section*{CRediT Authorship Contribution Statement}

\textbf{Luyao Xu:} Conceptualization, Data curation, Validation, Writing -review \& editing.
\textbf{Xiang Chen:} Software, Conceptualization, Methodology, Writing -review \& editing, Supervision.

\section*{Declaration of Competing Interests}
The authors declare that they have no known competing financial interests or personal relationships that could have appeared to influence the work reported in this paper.

\section*{Acknowledgement}

This research was partially supported by the National Natural Science Foundation of China (Grant no. 61202006) and the Open Project of State Key Laboratory for Novel Software Technology at Nanjing University under (Grant No. KFKT2024B21).

\section*{Data availability}
Data will be made available on request.

\bibliography{mylib}
\bibliographystyle{elsarticle}

\par
\vspace{1cm}
\noindent\textbf{Luyao Xu} is currently pursuing his Master degree at the School of Artificial Intelligence and Computer Science, Nantong University. His research interests include autonomous agent framework security. 

\vspace{1cm}
\noindent\textbf{Xiang Chen} received the B.Sc. degree in the school of management from Xi'an Jiaotong University, China in 2002. Then he received his M.Sc., and Ph.D. degrees in computer software and theory from Nanjing University, China in 2008 and 2011 respectively. He is currently an Associate Professor at the School of Artificial Intelligence and Computer Science, Nantong University. He has authored or co-authored more than 170 papers in refereed journals or conferences, such as IEEE Transactions on Software Engineering, ACM Transactions on Software Engineering and Methodology, Computer \& Security, IEEE Transactions on Reliability, Empirical Software Engineering, Information and Software Technology, Journal of Systems and Software, Software Testing, Verification and Reliability, Journal of Software: Evolution and Process, Automated Software Engineering, Software - Practice and Experience, Science of Computer Programming, Knowledge-based Systems, Engineering Applications of Artificial Intelligence, International Conference on Software Engineering (ICSE), International Conference on the Foundations of Software Engineering (FSE), International Conference Automated Software Engineering (ASE), International Symposium on Software Testing and Analysis (ISSTA), International Conference on Software Maintenance and Evolution (ICSME), International Conference on Program Comprehension (ICPC), International Symposium on Software Reliability Engineering (ISSRE) and International Conference on Software Analysis, Evolution and Reengineering (SANER). His research interests include software engineering, in particular software testing and maintenance, security vulnerability detection and understanding, large language models for software engineering, software repository mining, and empirical software engineering. He received two ACM SIGSOFT distinguished paper awards in ICSE 2021 and ICPC 2023. He is an editorial board member of Information and Software Technology. More information can be found at:
\url{https://xchencs.github.io}

\end{document}